\documentclass[runningheads]{llncs}
\usepackage[T1]{fontenc}
\usepackage{booktabs}
\usepackage{graphicx}
\usepackage[nolist]{acronym}
\usepackage{multirow}
\usepackage{amssymb}
\usepackage{mathtools}
\usepackage{comment}
\usepackage{siunitx}
\usepackage{pgfplots}
\usepackage{pgfplotstable}
\usepackage{lscape}
\usepackage{caption}
\usepackage{subcaption}
\usepackage[normalem]{ulem}
\robustify\uline
\usepackage{bm}
\usepackage{orcidlink}

\definecolor{tb_color_1}{RGB}{245,124,0}
\definecolor{tb_color_2}{RGB}{0,167,247}
\definecolor{tb_color_3}{RGB}{0,190,0}
\begin{document}
\begin{acronym}
        \acro{cow}[CoW]{Circle of Willis}
        \acro{topcow}[TopCoW]{Topolgy-Aware Anatomical Segmentation of the CoW}
        \acro{acdc}[ACDC]{Automated Cardiac Diagnosis Challenge}
        \acro{mra}[MRA]{Magnetic Resonance Angiography}
        \acro{cmr}[CMR]{Cardiac Magnetic Resonance}
        \acro{em}[EM]{electron microscopy}
        \acro{cldice}[clDice]{centerline Dice}
\end{acronym}
\title{Topologically Faithful Multi-class Segmentation in Medical Images}
\author{
Alexander H. Berger\inst{1, 4}\orcidlink{0009-0004-8843-7684} \and
Laurin Lux\inst{1,3}\orcidlink{0009-0003-7359-6212} \and
Nico Stucki\inst{1, 2, 3}\orcidlink{0009-0005-9736-9895} \and
Vincent Bürgin\inst{1}\orcidlink{0000-0002-8563-4189} \and
Suprosanna Shit\inst{1,6}\orcidlink{0000-0003-4435-7207} \and
Anna Banaszak\inst{1} \and
Daniel Rueckert\inst{1, 2, 3, 4, 5}\orcidlink{0000-0002-5683-5889} \and
Ulrich Bauer \inst{1, 2, 3}\orcidlink{0000-0002-9683-0724} \and
Johannes C. Paetzold\inst{5}\orcidlink{0000-0002-4844-6955}}

%index{Berger, Alexander H.} 
%index{Lux, Laurin}
%index{Stucki, Nico}
%index{Bürgin, Vincent}
%index{Shit, Suprosanna}
%index{Banaszak, Anna}
%index{Rueckert, Daniel}
%index{Bauer, Ulrich}
%index{Paetzold, Johannes C.}

%
\authorrunning{A. H. Berger et al.}
\institute{School of Computation, Information and Technology, Technical University of
Munich, Munich, Germany \and
Munich Data Science Institute, Technical University of Munich, Munich, Germany \and
Munich Center for Machine Learning, Munich, Germany \and
School of Medicine and Health, Klinikum rechts der Isar, Technical University of Munich, Munich, Germany  \and
Department of Computing, Imperial College London, London, UK \and
Department of Quantitative Biomedicine, University of Zurich, Switzerland
}
\maketitle              % typeset the header of the contribution
\begin{abstract}
Topological accuracy in medical image segmentation is a highly important property for downstream applications such as network analysis and flow modeling in vessels or cell counting. Recently, significant methodological advancements have brought well-founded concepts from algebraic topology to binary segmentation. However, these approaches have been underexplored in multi-class segmentation scenarios, where topological errors are common. We propose a general loss function for topologically faithful multi-class segmentation extending the recent Betti matching concept, which is based on induced matchings of persistence barcodes. We project the $N$-class segmentation problem to $N$ single-class segmentation tasks, which allows us to use 1-parameter persistent homology, making training of neural networks computationally feasible.
We validate our method on a comprehensive set of four medical datasets with highly variant topological characteristics. Our loss formulation significantly enhances topological correctness in cardiac, cell, artery-vein, and Circle of Willis segmentation.\footnote{The code is available at github.com/AlexanderHBerger/multiclass-BettiMatching}

\end{abstract}

\section{Introduction}
Topological correctness is crucial for many downstream tasks, such as blood flow modeling or cell counting. Modern segmentation networks achieve high pixel-wise accuracy but often cannot preserve important topological features \cite{bentaieb2016topology,hu2019topology}. Recent works addressed this issue via a variety of methods such as postprocessing \cite{byrne2022persistent,li2023robust}, injecting topological priors during training \cite{byrne2022persistent,clough2020topological,clough2019explicit,gupta2022learning}, or topology-aware loss functions \cite{hu2021discretemorset,hu2021warping,hu2019topology,qiu2023corsegrec,shit2021cldice,stucki2023topological}.
These methods come with varying degrees of generalization capabilities and theoretical guarantees. For example, some persistent homology-based losses cannot guarantee the spatial matching of topological features \cite{clough2020topological,hu2019topology}. Recently, the \emph{Betti matching} loss was proposed as a general loss function to train binary segmentation networks with strict topological guarantees and correct spatial alignment of topological features \cite{stucki2023topological}. Despite these successes, the realm of topology-preserving multi-class segmentation has been under-explored. In this work, we fill this gap by proposing a generalizable multi-class segmentation loss extending the Betti matching concept to multi-class settings. Based on popular methods, specifically, clDice \cite{shit2021cldice}, and HuTopo \cite{hu2019topology}, we establish other topology-aware multi-class baselines. Furthermore, we introduce a weighting term based on topological structures that can improve the topological correctness depending on the dataset characteristics. In experiments on four topologically variant datasets, we demonstrate our method's utility and outperform all baselines.

\begin{figure}[t]
    \centering
    \includegraphics[width=1\linewidth]{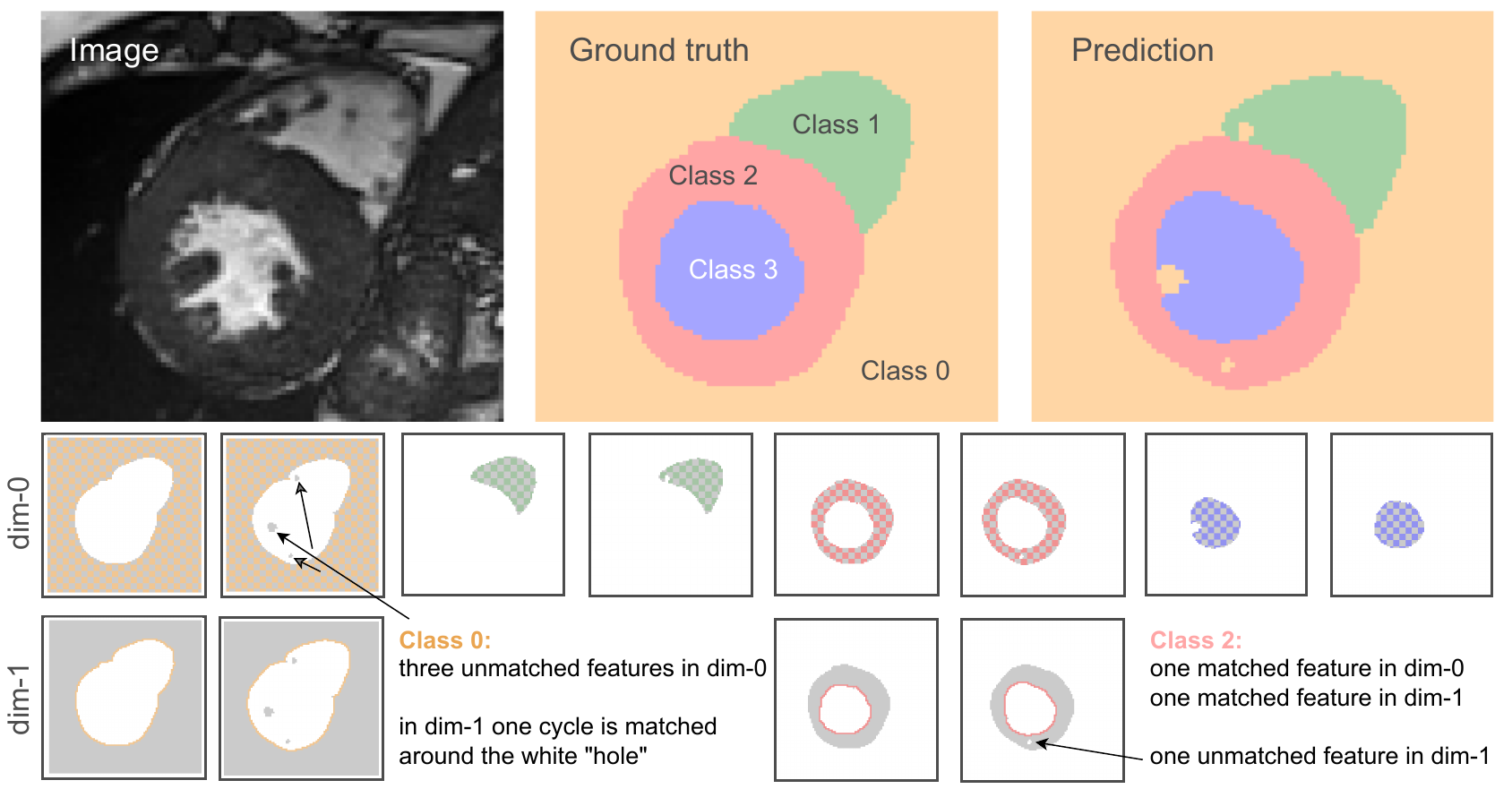}
    \caption{Top: image, ground truth, and exemplary segmentation of the cardiac dataset. Bottom: ground truth (left) next to the segmentation (right), pairwise for every class. With our multi-class Betti-matching formulation, we match connected components (dim-0) and cycles (dim-1) in each individual class. Matched features in dim-0 and dim-1 are colored in a checkerboard pattern and have colored feature cycles, respectively. All classes' matched and unmatched features guide our loss function for multi-class topology-preserving segmentation.}
    \label{fig:enter-label}
\end{figure}
\subsubsection{Related works}
A plethora of methods have emphasized the importance of topologically correct segmentation in medical imaging. These methods can be assigned to three categories. Firstly, post-processing frameworks that aim to fix topological errors of preliminary segmentations \cite{li2023robust,mosinska2018beyond,wang2018post}. These approaches do not integrate with end-to-end training of neural networks. The second category uses fixed topological priors during training or post-processing \cite{byrne2022persistent}. Clough et al. \cite{clough2020topological} follow an approach where they defined the expected topology using Betti numbers on which they computed a loss term. Gupta et al. \cite{gupta2022learning} apply physiological priors to segment various medical structures. Other studies use similar approaches for coronary artery segmentation \cite{zhang2022progressive,zhang2023topology}. However, these methods do not generalize to other tasks and are only applicable when every sample for the segmentation task has the same topological features. The last category contains topologically faithful loss functions in which a topological loss is computed without using task-specific knowledge. These methods are often based on skeletonization for tubular structures \cite{shit2021cldice,qiu2023corsegrec,lin2023structure}. In a foundational work, Hu et al. \cite{hu2019topology} proposed using a loss function for image segmentation, which minimizes the squared distance of matched points in persistence diagrams of dimension 1. However, this method does not guarantee that matched structures are spatially related in any sense, which has a significant negative impact on training \cite{stucki2023topological}. A recent work overcame the limitations of these methods and showed that induced matchings can achieve a spatially correct matching between barcodes in a binary segmentation setting with formal guarantees \cite{stucki2023topological}.
Other approaches applied homotopy warping to identify critical pixels and measure topological differences between images \cite{hu2021warping,jain2010boundary} or utilized discrete Morse theory to compare critical topological structures \cite{hu2021discretemorset}. Such loss functions are beneficial because they do not require prior knowledge, they can be generalized across tasks and can be incorporated into end-to-end training. However, research regarding such general, homology-based loss functions in multi-class settings has been limited.

\section{Method}
\label{sec:method}
This work presents a loss formulation that captures topological errors in multi-class segmentation masks. Our loss can be used to train arbitrary segmentation networks in an end-to-end fashion. We utilize persistent homology to capture topological features at multiple scales and generalize homology-based loss functions to multi-class segmentations. 

\subsubsection{Overview of persistent homology for image segmentation.}
Let $x \in \mathbb{R}^{W \times H}$ be a grayscale image, $y \in \{0,1\}^{N \times W \times H}$ its segmentation in $N$ mutually exclusive classes, and $\hat{y} \in [0,1]^{N \times W \times H}$ a predicted likelihood map for each class with $\sum_{c=1}^{N} \hat{y}_{c,i,j} = 1$. 
In a binary segmentation setting, the predicted likelihood map $\hat{y} \in [0,1]^{W \times H}$ can be seen as a function $\hat{y} \colon K^{W \times H} \to [0,1]$ on a \emph{cubical grid complex} $K^{W \times H}$ (see \cite{stucki2023topological}) and its sublevel filtration $K_t = \hat{y}^{-1}((-\infty,t])$ as a model for its topology. Since the sublevel filtration depends only on one parameter $t \in \mathbb{R}$, the persistent homology of this filtration forms a 1-parameter persistence module, and its barcode decomposition can be used as a descriptor for the topology of the prediction. For a thorough theoretical description of persistent homology for binary image segmentation, we refer the reader to \cite{stucki2023topological}.

\subsubsection{Single-class loss functions for multi-class segmentations.}
In a multi-class segmentation task, the predicted likelihood map $\hat{y} \in [0,1]^{N \times W \times H} \to [0,1]$ corresponds to a function $\hat{y} \colon K^{W \times H} \to [0,1]^N$. Since its sublevel filtration $K_{(t_n)} = \{x \in K^{W \times H} \mid \hat{y}(x) \leq (t_1,\ldots,t_N)\}$ depends on $N$ parameters $t_1,\ldots,t_N \in \mathbb{R}$, its persistent homology forms an $N$-parameter persistence module. The complexity of computing descriptors for $N$-parameter persistent homology makes it infeasible to train segmentation networks with such a setup.

In order to circumvent this problem, consider a multi-class segmentation network $f \colon \mathbb{R}^{W \times H} \to [0,1]^{N \times W \times H}$ and denote by 
\begin{equation}
p_c \colon [0,1]^{N \times W \times H} \to [0,1]^{W \times H}, \ (\hat{y}_{n,i,j}) \mapsto (\hat{y}_{c,i,j})
\end{equation}
the restriction to the $c$-th channel. Then $f_c \vcentcolon = p_c \circ f$ can be seen as the solution to a single-class segmentation task predicting the likelihood of a pixel being within class $c$. This way, we can break down the multi-class segmentation task with $N$ classes into $N$ single-class segmentation tasks, allowing us to apply single-class segmentation loss functions. Thus, any single-class segmentation loss $\mathcal{L} \colon [0,1]^{W\times H} \times \{0,1\}^{W\times H} \to \mathbb{R}$ can be extended to a multi-class segmentation loss by summing the individual losses of each channel:
\begin{equation}
    \mathcal{L} \colon [0,1]^{N \times W \times H} \times \{0,1\}^{N \times W \times H} \to \mathbb{R}, \ (\hat{y},y) \mapsto \sum_{c=1}^N \mathcal{L}(p_c(\hat{y}),p_c(y))
\end{equation}

\noindent Considering each channel $c$ individually and computing the barcode of its sublevel filtration corresponds to slicing the $N$-parameter persistence module along the line
\begin{equation}
\{t \in \mathbb{R}^N \mid t_c = t, t_n = 1-t \text{ for } n \neq c\}
\end{equation}
to a $1$-parameter module and computing its corresponding barcode decomposition, which scales the computational complexity linearly with the number of classes $N$ compared to a binary setting. Therefore, the total computational complexity is dominated by the barcode computation that scales cubically with the number of pixels \cite{stucki2023topological}. We utilize the efficient implementation for barcode computation presented in \cite{stucki2024efficient} and provide runtime information in the supplement.
Notably, the presented formulation naturally extends to other topology-preserving losses, which allows us to extend and implement other binary topology-preserving losses to our setting and compare their generalizability, which we do in the experiment section. Furthermore, they extend to topological metrics, especially the Betti matching error \cite{stucki2023topological}.

\subsubsection{Weighting.}
The Betti matching loss $l_{BM}$ is based on the induced matching of persistence barcodes $\mu(\bm{L},\bm{G})$ (as described in \cite{stucki2023topological}), which consists of matched and unmatched features between ground truth and prediction (as depicted in Fig. \ref{fig:enter-label}). Therefore, $l_{BM}$ can be decomposed into one loss component for matched $l_{BM}^{\text{m}}$ and one for unmatched $l_{BM}^{\text{u}}$ features, respectively. $l_{BM}^{\text{m}}$ reinforces topological features that are correctly predicted by the network, and $l_{BM}^{\text{u}}$ penalizes topological features that are predicted by the network but do not exist in the ground truth. In \cite{stucki2023topological}, these components are implicitly weighted equally. However, in a multi-class setting, with varying numbers of classes and topological features, one loss component can dominate the other, which may deteriorate performance (see Sec. \ref{sec:abl_weights}). Hence, we propose a weighting of these components with the parameters $\gamma^{\text{m}}$ and $\gamma^{\text{u}}$. 
%Also, note that homology-based loss functions do not suffer from pixel-wise class imbalances, which is a common problem for pixel-based loss functions \cite{ma2021loss} in biomedical multi-class segmentation. 
Lastly, we combine our final multi-class topological loss with a standard pixel-based loss function (e.g., Dice) to the final loss function:
\begin{equation}
    l_{\text{total}} = \alpha \cdot (\gamma^{\text{m}} \cdot l_{BM}^{\text{m}} + \gamma^{\text{u}} \cdot l_{BM}^{\text{u}}) + l_{\text{Dice}}
\end{equation}

\section{Data and Experimentation}
\subsubsection{Datasets.}We evaluate our method on four different, public multi-class segmentation tasks. While all data is from different modalities, we choose tasks where topological correctness is a major objective (especially for downstream applications) and has proven to be challenging with traditional methods. For training/validation/test splits and further details, please see the Supplement.

First, in the \ac{acdc} challenge dataset \cite{bernard2018acdc}, the task is cardiac segmentation, i.e., segmenting the myocardium and the left and right ventricles in 2D short-axis slices of \ac{cmr} scans. Here, pathological samples often lead to failure modes that exhibit topological errors \cite{byrne2022persistent}. We use each slice, including the apex and base, to get data points with varying topological features.
%The 150 patient scans' vary between 154x154x8 and 256x256x10 voxels. We randomly crop slices of size 154x154 around the left ventricle and use each slice as a separate sample, including the apex and base, to get data points with varying topological features. We use the original data split with 100 training samples (exams) and 50 testing samples. 
%
Second, we use a publicly available cellular and subcellular \ac{em} dataset (Platelet) with six classes (cell, mitochondrion, canalicular channel, alpha granule, dense granule, and dense granule core) \cite{guay2021platelet}.
%that contains 50 samples for training and 25 for testing each (800$\times$800 pixels) with six classes (cell, mitochondrion, canalicular channel, alpha granule, dense granule, and dense granule core) \cite{guay2021platelet}. We create overlapping patches of size 200$\times$200 during our experiments.
%
%
Third, we test our method on artery-vein classification in OCTA images with the OCTA-500 dataset \cite{li2024octa}.
%that contains 200 en-face OCTA images with a field of view of 3mm$\times$3mm and a patch size of 301x301. We randomly select 40 scans for testing.  
%
Last, we test our model on \ac{cow} segmentation in \ac{mra} scans with 15 vessel components as classes. The \ac{cow} has hypoplastic and absent components across different subjects, making correct segmentation challenging \cite{yang2023topcow}. We project the \ac{mra} scan and the label to a 2D image and segmentation mask.
%use the public MICCAI 2023 \ac{topcow} challenge data consisting of 110 subjects and randomly select 22 subjects for testing. For each subject, we project the \ac{mra} scan and the label to a 2D image and 2D segmentation mask, which is scaled to 100x80 pixels. 

\subsubsection{Baselines.}
We employ three baselines against which we evaluate the performance of our loss formulation. All baselines enable end-to-end training and are independent of the used network. First, the generalized Dice loss \cite{sudre2017generalised}, which is still one of the most frequently used loss terms in biomedical image segmentation due to its inherent class balancing properties \cite{ma2021loss}. Thus, we choose the Dice loss as a standalone baseline. Second, we combine the Dice loss with the (multi-class extended) \ac{cldice} loss \cite{shit2021cldice}. \ac{cldice} is especially effective for segmenting tubular structures. %and gives strong topological guarantees for their topology. 
Lastly, we develop an additional baseline where we combine the Dice loss with a multi-class extension of the HuTopo loss \cite{hu2019topology} via the multiclass generalization introduced in Section \ref{sec:method}.

\subsubsection{Metrics}
We evaluate segmentation performance with a comprehensive set of topology and pixel-wise scores. Specifically: the \textit{Betti matching error}, \textit{Betti number errors}, \textit{Dice}, and \textit{clDice}. The Betti matching error (BM) is the most accurate indicator of faithful topological segmentations \cite{stucki2023topological}.
%- potentially ARI, haus
All metrics are averaged across all foreground classes without weighting.

\subsubsection{Training and model selection}
We train a U-Net architecture \cite{ronneberger2015u} with residual units from scratch. We use the Adam optimizer \cite{kingma2014adam}, a fixed learning rate, and sigmoid scheduling for $\alpha$ (see Supplement). Note that our loss formulation is independent of the underlying network architecture. We perform 5-fold cross-validation and evaluate on an independent test set. We perform a random hyperparameter search with 50 runs on each of the splits and select the model that has the highest performance $S$ on the validation set with $S$ being defined as a balanced performance metric of pixel-wise accuracy and topological faithfulness:

\begin{equation}
    S = \operatorname{Dice} + \bigg(1 - \min\bigg(1, \frac{l_{BM}}{\beta_0 + \beta_1}\bigg)\bigg)
\end{equation}

\noindent where $\beta_i$ is the Betti number in dimension $i$. We report the mean performance and standard deviation on the independent test set across the five data splits. Please refer to the supplement for hyperparameters. We use the paired t-test between our model and each baseline to evaluate statistically significant performance ($p$-value $<0.05$) improvements.

\section{Results}

\begin{table}[t]
    \centering

    \caption{Quantitative results. We show the performance of our multiclass Betti matching loss against multiple other topology-aware losses that we adapted to the multiclass segmentation setting. Best performances indicated in \textbf{bold}, statistical significance \underline{underlined} ($p$-value $<0.05$).}
    \label{tab:results}
    \begin{tabular}{llccccc}
    \toprule
    \textbf{Dataset}                               & \multicolumn{1}{l}{\textbf{Loss}} & \textbf{Dice $\uparrow$} & \textbf{clDice $\uparrow$}& \textbf{BM $\downarrow$} & \textbf{B0 $\downarrow$} & \textbf{B1 $\downarrow$} \\ \midrule
    \multirow{4}{*}{ACDC}                          & Dice                               & $.868{\scriptstyle \pm.023}$& \uline{$.636{\scriptstyle \pm.004}$}& \uline{$0.130{\scriptstyle \pm0.014}$}& \uline{$0.094{\scriptstyle \pm0.012}$}& \uline{$0.032{\scriptstyle \pm0.006}$}\\
                                                   & ClDice                             & $\bm{.871}{\scriptstyle \pm\bm{.020}}$& \uline{$.629{\scriptstyle \pm.009}$}& \uline{$0.380{\scriptstyle \pm0.226}$}& \uline{$0.115{\scriptstyle \pm0.014}$}& \uline{$0.261{\scriptstyle \pm0.217}$}\\
                                                   & HuTopo                        & $.862{\scriptstyle \pm.005}$& $.634{\scriptstyle \pm.004}$& \uline{$0.219{\scriptstyle \pm0.049}$}& \uline{$0.164{\scriptstyle \pm0.044}$}& \uline{$0.049{\scriptstyle \pm0.010}$}\\
                                                   & Ours                               & $.862{\scriptstyle \pm.008}$& $\bm{.641}{\scriptstyle \pm\bm{.005}}$& $\bm{0.064}{\scriptstyle \pm\bm{0.007}}$& $\bm{0.038}{\scriptstyle \pm\bm{0.005}}$& $\bm{0.022}{\scriptstyle \pm\bm{0.004}}$\\ \midrule
    \multicolumn{1}{l}{\multirow{4}{*}{Platelet}} & Dice                               & $.685{\scriptstyle \pm.015}$& \uline{$.548{\scriptstyle \pm.009}$}& \uline{$1.289{\scriptstyle \pm0.147}$}& \uline{$0.643{\scriptstyle \pm0.072}$}& \uline{$0.280{\scriptstyle \pm0.045}$}\\
    \multicolumn{1}{l}{}                          & ClDice                             & $.682 {\scriptstyle \pm.016 }$& \uline{$.544{\scriptstyle \pm.011}$}& \uline{$4.682{\scriptstyle \pm2.800}$}& \uline{$1.053{\scriptstyle \pm0.293}$}& \uline{$3.238{\scriptstyle \pm2.474}$}\\
    \multicolumn{1}{l}{}                          & HuTopo                        & \uline{$.635{\scriptstyle \pm.042}$}& \uline{$.517{\scriptstyle \pm.018}$}& \uline{$1.420{\scriptstyle \pm0.408}$}& \uline{$0.684{\scriptstyle \pm0.218}$}& $0.301{\scriptstyle \pm0.136}$       \\
    \multicolumn{1}{l}{}                          & Ours                               & $\bm{.696}{\scriptstyle \pm\bm{.019}}$& $\bm{.564}{\scriptstyle \pm\bm{.012}}$& $\bm{0.978}{\scriptstyle \pm\bm{0.126}}$& $\bm{0.483}{\scriptstyle \pm\bm{0.068}}$& $\bm{0.191}{\scriptstyle \pm\bm{0.024}}$\\ \midrule
    \multirow{4}{*}{OCTA-500}& Dice                               & $\bm{.829}{\scriptstyle \pm\bm{.006}}$& $\bm{.567}{\scriptstyle \pm\bm{.004}}$& \uline{$34.258{\scriptstyle \pm2.929}$}& $11.825{\scriptstyle \pm2.518}$& $0.018{\scriptstyle \pm0.019}$\\
                                                   & ClDice                             & $.794{\scriptstyle \pm.013}$& $.560{\scriptstyle \pm.005}$& \uline{$23.510{\scriptstyle \pm0.498}$}& $\bm{6.942}{\scriptstyle \pm\bm{0.367}}$& $\bm{0.008}{\scriptstyle \pm\bm{0.011}}$\\
                                                   & HuTopo                        & ${.787\scriptstyle \pm.028}$& ${.554\scriptstyle \pm.009}$& \uline{${28.285\scriptstyle \pm3.871}$}& ${8.255\scriptstyle \pm1.877}$& ${0.035\scriptstyle \pm0.021}$\\
                                                   & Ours                               & $.798{\scriptstyle \pm.013}$& $.556{\scriptstyle \pm.007}$& $\bm{17.950}{\scriptstyle \pm\bm{0.567}}$& $9.670{\scriptstyle \pm0.667}$& $0.055{\scriptstyle \pm0.021}$\\ \midrule
    \multirow{4}{*}{TopCoW}                        & Dice                               & $.625{\scriptstyle \pm.148}$& $.624{\scriptstyle \pm.145}$& \uline{$1.791{\scriptstyle \pm1.176}$}& \uline{$1.495{\scriptstyle \pm1.114}$}& $0.042{\scriptstyle \pm0.021}$\\
                                                   & ClDice                             & $.685{\scriptstyle \pm.102}$& $.693{\scriptstyle \pm.109}$& $1.043{\scriptstyle \pm0.803}$& $0.835{\scriptstyle \pm0.713}$& $0.032{\scriptstyle \pm0.017}$\\
                                                   & HuTopo                        & $.670{\scriptstyle \pm.138}$& $.675{\scriptstyle \pm.141}$& $1.442{\scriptstyle \pm1.668}$& $1.207{\scriptstyle \pm1.585}$& $0.042{\scriptstyle \pm0.025}$\\
                                                   & Ours                               & $\bm{.717}{\scriptstyle \pm\bm{.021}}$& $\bm{.725}{\scriptstyle \pm\bm{.025}}$& $\bm{0.773}{\scriptstyle \pm\bm{0.304}}$& $\bm{0.616}{\scriptstyle \pm\bm{0.231}}$& $\bm{0.027}{\scriptstyle \pm\bm{0.017}}$\\ \bottomrule
    \end{tabular}
\end{table}
\begin{figure}[t]
    \centering
    \includegraphics[width=\linewidth]{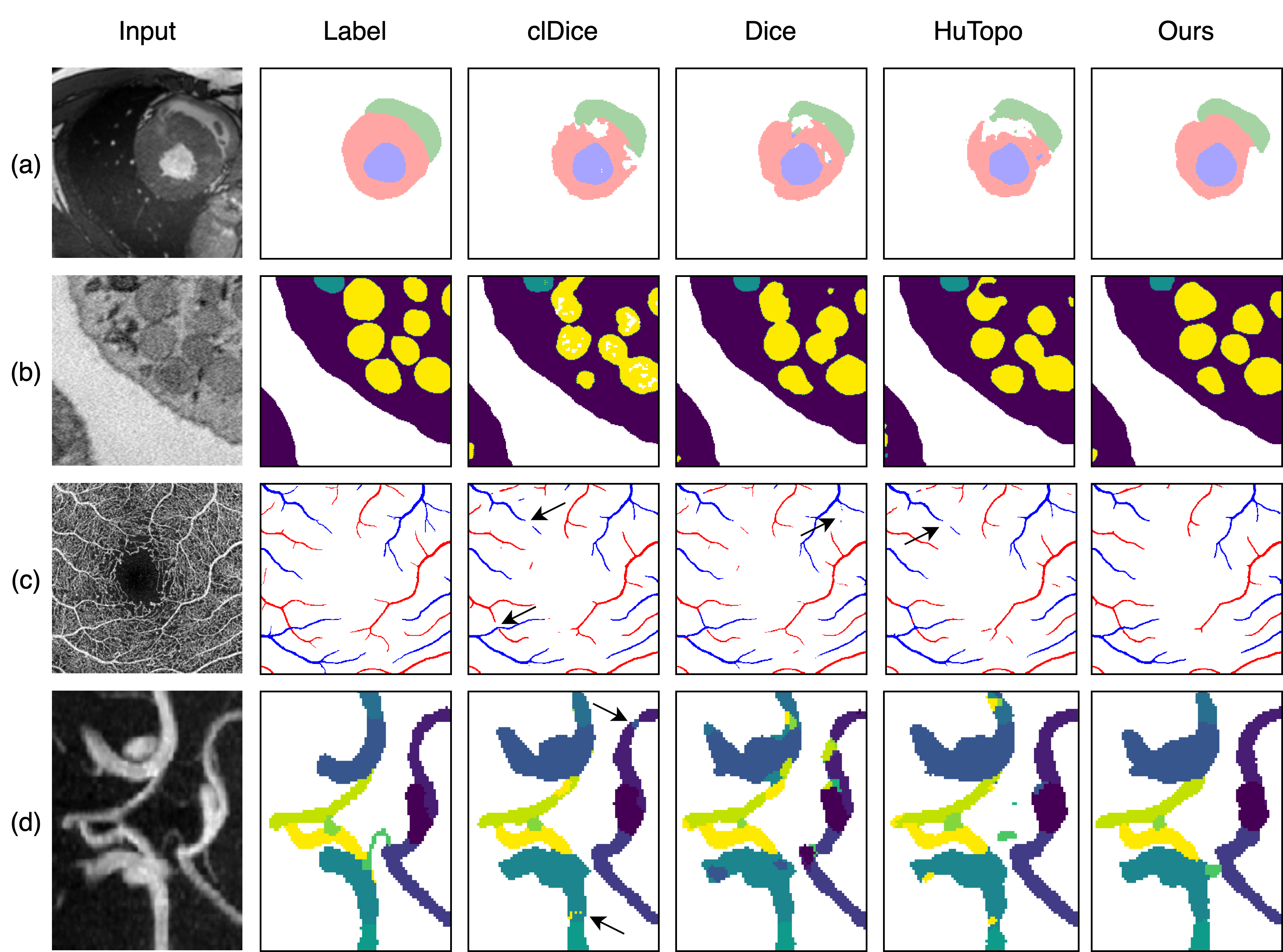}
    \caption{Qualitative results on ACDC (a), Platelet (b), OCTA-500 (c), and TopCoW (d) dataset. Our method improves topological correctness in all multi-class segmentation tasks. We indicate some topological errors with black arrows.}
    \label{fig:qualitatives}
\end{figure}

Our results comprehensively demonstrate that our proposed multi-class segmentation loss outperforms all baselines across all datasets in Betti matching errors. Further, we outperform all baselines in Betti number errors 0 and 1 as well as clDice in the Platelet, TopCoW, and ACDC datasets; see Table \ref{tab:results}.
Even in the non-topology-aware Dice score, our method performs on par with the baselines across most datasets. 
Overall, we find that the (multi-class extended) HuTopo baseline \cite{hu2019topology} does not generalize well to the multi-class setting. We attribute this to an amplification of the possible incorrect matchings of persistence features as described in Stucki et al. \cite{stucki2023topological}; i.e., the matched features do not correspond spatially. 
Further, we find specific results for the individual datasets:

%\noindent \textit{\textbf{ACDC.}}
\subsubsection{ACDC.}
This dataset exhibits a mostly regular structure where the three classes have one unique topology, i.e., the Myocardium forms one connected component and mostly one cycle. Achieving good volumetric performance is rather simple; all methods show similar, high performance in Dice and clDice; however, our method still significantly improves the Betti matching error and the Betti number errors. Our method is especially beneficial in the irregular (e.g., basal) slices. There, the structures do not have their typical topology, making approaches based on topological post-processing unsuitable.
%\noindent \textit{\textbf{Platelet.}} 
\subsubsection{Platelet.}
In this dataset, the aim is to segment round objects where the topology is described by "inclusion," e.g., a mitochondrion is always inside a cell segment. Our method is superior to all baselines across all metrics. The HuTopo and clDice baselines perform especially poorly, which we attribute to the large image size (200$\times$200 pixels), making HuTopo's spatial mismatch more prevalent, and the lack of tubular structures, negatively impacting clDice performance. Qualitatively, we observe that the baselines often merge individual structures.
%\noindent \textit{\textbf{OCTA-500.}} 
\subsubsection{OCTA-500.}
Here, the Dice loss results in the best Dice score, which is in contrast to the other experiments. We hypothesize that the Dice loss is more robust against the dataset's annotation scarcity \cite{menten2023synthetic,kreitner2024synthetic}. However, our loss significantly decreases the BM error compared to all baselines, including the Dice loss. The qualitative examples show that all baselines suffer from a large number of discontinuous vessel segments, which can be critical for downstream applications.
%\noindent \textit{\textbf{TopCoW.}}
\subsubsection{TopCoW.}
Our method improves performance across all metrics. This dataset is challenging because of its 13 distinct classes, which partially overlap in the 2D projection. In this setting, all topology-aware losses perform more robustly than the Dice loss. We conclude that in such challenging settings, topological methods are especially useful. Furthermore, our method shows by far the lowest standard deviation, further emphasizing its robustness compared to other methods.

\subsubsection{Ablation on weighting.}
\label{sec:abl_weights}
\begin{figure}[t]
    \centering
    \begin{subfigure}{0.45\textwidth}
        \centering
        \begin{tikzpicture}
            %\pgfplotsset{set layers}
            \begin{axis}[
                grid=none,
                grid style={solid,gray!30!white},
                axis lines=middle,
                xlabel={$\gamma^{\text{m}}$},
                ylabel={BM Error},
                x label style={at={(axis description cs:0.5,-0.1)},anchor=north},
                y label style={at={(axis description cs:-0.18,.5)},rotate=90,anchor=south},
                axis y line*=left, % Use left y-axis
                ymin=18, ymax=34,
                width=\linewidth, height=4.3cm
            ]
            \addplot[only marks, mark=*, color=tb_color_2] table [x=weight_matched, y=bm_loss, col sep=comma] {tables/abl_weight.csv};
            \addplot[line width=2pt, color=tb_color_2, smooth] table [x=weight_matched, y={create col/linear regression={y=bm_loss}}, col sep=comma] {tables/abl_weight.csv};
            \end{axis}
            %\node at (current axis.above origin) [anchor=south, yshift=-4.1cm, xshift=2.25cm] {(a)};
        \end{tikzpicture}
        %\caption{Betti Matching error with varying $\gamma_{BM}^{\text{m}}$}
        \label{fig:abl_wighting_a}
    \end{subfigure}
    \hfill
    \begin{subfigure}{0.45\textwidth}
        \centering
        \begin{tikzpicture}
            %\pgfplotsset{set layers}
            \begin{axis}[
                grid=none,
                grid style={solid,gray!30!white},
                axis lines=middle,
                xlabel={$\gamma^{\text{m}}$},
                ylabel={Dice Score},
                x label style={at={(axis description cs:0.5,-0.1)},anchor=north},
                y label style={at={(axis description cs:-0.18,.5)},rotate=90,anchor=south},
                axis y line*=left, % Use right y-axis
                ymin=0.73, ymax=0.88,
                width=\linewidth, height=4.3cm
            ]
            \addplot[only marks, mark=*, color=gray, opacity=0.5] table [x=weight_matched, y=dice, col sep=comma] {tables/abl_weight.csv};
            \addplot[line width=2pt, color=gray, opacity=0.5, smooth] table [x=weight_matched, y={create col/linear regression={y=dice}}, col sep=comma] {tables/abl_weight.csv};
            \end{axis}
            %\node at (current axis.above origin) [anchor=south, yshift=-4.1cm, xshift=2.25cm] {(b)};
        \end{tikzpicture}
        %\caption{}
        \label{fig:abl_wighting_b}
    \end{subfigure}
    \caption{Betti Matching error (left) and Dice score (right) with varying $\gamma^{\text{m}}$}
    \label{fig:abl_wighting}
\end{figure}
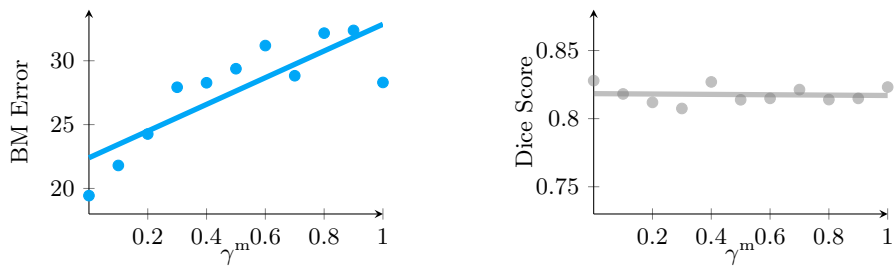

In an ablation, we study the effect of our proposed weighting concept, see Section \ref{sec:method}. We find that topological weighting can strongly affect the model's performance regarding the Betti matching error while being robust in terms of pixel-wise accuracy, see Fig. \ref{fig:abl_wighting}. Here, we study the OCTA-500 dataset and show that a low weight for the loss of matched topological features $l_{BM}^{\text{m}}$ (and thereby a relatively higher weight for the loss of unmatched topological features $l_{BM}^{\text{u}}$) drastically improves topological performance while having a negligible effect on Dice. We attribute this observation to the predominance of topological features in dimension 0 in the hierarchically structured vasculatures in OCTA images. Note that other datasets do not follow this trend indicating that the weighting parameter must be specifically tuned for the used dataset.

%and propose its study in further research. 

\section{Conclusion}
Our study introduces a novel loss for multi-class image segmentation, extending the Betti matching concept to preserve topology. By decomposing the $N$-class problem, we circumvent the use of multi-parameter persistent homology, facilitating neural network training. Our empirical findings are three-fold. First, we find that our proposed generalization can successfully extend topology-aware losses to multiclass problems. Second, we provide strong empirical evidence that our loss does not impede pixel-wise accuracy and significantly improves topological accuracy. Finally, we find that our loss function consistently outperforms other topology-preserving baselines across several datasets, rendering it suitable for topologically critical applications. 

\begin{credits}
\subsubsection{\ackname} A. Berger is supported by the Stiftung der Deutschen Wirtschaft.

\subsubsection{\discintname}
The authors have no competing interests to declare that are relevant to the content of this article.
\end{credits}
\bibliographystyle{splncs04}
\bibliography{mybib}

\clearpage
\section{Supplementary Material}
\subsection*{Alpha scheduling}
\begin{equation}
    \alpha = \left(\frac{2}{1 + e^{-10p}} - 1\right) \cdot \alpha_{\mathrm{max}}
\end{equation}
with: 
\begin{equation}
p = \frac{{\text{{step}} - \text{warmup}_{\alpha}}}{{\text{{total\_steps}}}}
\end{equation}
\begin{table}[h]
\centering
\label{tab:datasets}
\caption{Overview of the used datasets, the respective image sizes, and the train, validation, and test splits.}
\begin{tabular}{lccccc}
\toprule
\textbf{Dataset}  & \textbf{Structures}      & \textbf{Patch-Size}        & \multicolumn{1}{c}{\textbf{\begin{tabular}[c]{@{}c@{}}Training \\ samples\end{tabular}}} & \multicolumn{1}{c}{\textbf{\begin{tabular}[c]{@{}c@{}}Validation \\ samples\end{tabular}}} & \multicolumn{1}{c}{\textbf{\begin{tabular}[c]{@{}c@{}}Test \\ samples\end{tabular}}} \\
\midrule
ACDC     & Cardiac         & $154 \times 154$  & 4596          & 1110               & 3228         \\
Platelet & (Sub-)Cellular  & $200 \times 200 $ & 1440          & 360                & 864          \\
OCTA-500 & Retinal         & $301 \times 301$  & 128           & 32                 & 40           \\
TopCoW   & Cerebrovascular & $100 \times 80$   & 70            & 18                 & 22           \\
\bottomrule
\end{tabular}
\end{table}

\begin{table}[h]
    \centering
    \caption{Hyperparameter search space for model hyperparameters.}
    \begin{tabular}{lcccccc}
    \toprule
    \textbf{Dataset}  & \multicolumn{1}{c}{\textbf{\begin{tabular}[c]{@{}c@{}}max \\ epochs\end{tabular}}} & \multicolumn{1}{c}{\textbf{\begin{tabular}[c]{@{}c@{}}learning rate \\ samples\end{tabular}}} & \multicolumn{1}{c}{\textbf{\begin{tabular}[c]{@{}c@{}}num \\ layers\end{tabular}}} & \multicolumn{1}{c}{\textbf{\begin{tabular}[c]{@{}c@{}}num res. \\ units\end{tabular}}} & \textbf{batch size} \\
    \midrule
    ACDC     & 100                                                  & $[0.0001,0.01]$                                          & $\{4,5\}$                                              & $\{2,3,4,5\}$                                                & $\{8,16,32,64,128\}$ \\
    Platelet & 100                                                  & $[0.00005,0.005]$                                        & $\{4,5\}$                                              & $\{2,3,4,5\}$                                                & $\{8,16,32,64,128\}$ \\
    OCTA-500 & 300                                                  & $[0.0001,0.01]$                                          & $\{4,5\}$                                              & $\{2,3,4,5\}$                                                & $\{8,16,32,64\}$ \\
    TopCoW   & 300                                                  & $[0.0001,0.01]$                                          & $\{4,5\}$                                              & $\{2,3,4,5\}$                                                & $\{8,16,32\}$ \\
    \bottomrule
    \end{tabular}
    \end{table}
    
    \begin{table}[h!]
    \centering
    \caption{Hyperparameter search space for loss-related hyperparameters.}
    \vspace{1em}
    \begin{tabular}{lcccccc}
    \toprule
    \textbf{Dataset}  & \textbf{$\alpha_{\mathrm{clDice}}$} & \textbf{$\alpha$}       & \textbf{$\text{warmup}_{\alpha}$} & \multicolumn{1}{c}{\textbf{\begin{tabular}[c]{@{}c@{}}ignore \\ background\end{tabular}}} \\
    \midrule
    ACDC     & $[0.1,0.8]$                & $[0.001,0.1]$  & $\{0,10,20,50\}$         & $\{\text{true},\text{false}\}$                              \\
    Platelet & $[0.1,0.8]$                & $[0.001,0.1]$  & $\{0,10,20,50\}$         & $\{\text{true},\text{false}\}$                              \\
    OCTA-500 & $[0.1,0.8]$                & $[0.001,0.05]$ & $\{0,20,50,100\}$        & $\{\text{true},\text{false}\}$                              \\
    TopCoW   & $[0.1,0.8]$                & $[0.001,0.1]$  & $\{0,10,20,50\}$        & $\{\text{true},\text{false}\}$                              \\
    \bottomrule
    \end{tabular}
    \end{table}

\begin{figure}[h]
    \centering
    \begin{subfigure}{0.45\textwidth}
        \centering
        \begin{tikzpicture}
            %\pgfplotsset{set layers}
            \begin{axis}[
                grid=none,
                grid style={solid,gray!30!white},
                axis lines=middle,
                xlabel={$\gamma^{\text{m}}$},
                ylabel={BM Error},
                x label style={at={(axis description cs:0.5,-0.1)},anchor=north},
                y label style={at={(axis description cs:-0.18,.5)},rotate=90,anchor=south},
                axis y line*=left, % Use left y-axis
                ymin=0.02, ymax=0.18,
                width=\linewidth, height=4.3cm
            ]
            \addplot[only marks, mark=*, color=tb_color_2] table [x=weight_matched, y=bm_loss, col sep=comma] {tables/abl_weight2.csv};
            \addplot[line width=2pt, color=tb_color_2, smooth] table [x=weight_matched, y={create col/linear regression={y=bm_loss}}, col sep=comma] {tables/abl_weight2.csv};
            \end{axis}
            %\node at (current axis.above origin) [anchor=south, yshift=-4.1cm, xshift=2.25cm] {(a)};
        \end{tikzpicture}
        %\caption{Betti Matching error with varying $\gamma_{BM}^{\text{m}}$}
        \label{fig:abl_wighting_a}
    \end{subfigure}
    \hfill
    \begin{subfigure}{0.45\textwidth}
        \centering
        \begin{tikzpicture}
            %\pgfplotsset{set layers}
            \begin{axis}[
                grid=none,
                grid style={solid,gray!30!white},
                axis lines=middle,
                xlabel={$\gamma^{\text{m}}$},
                ylabel={Dice Score},
                x label style={at={(axis description cs:0.5,-0.1)},anchor=north},
                y label style={at={(axis description cs:-0.18,.5)},rotate=90,anchor=south},
                axis y line*=left, % Use right y-axis
                ymin=0.8, ymax=0.9,
                width=\linewidth, height=4.3cm
            ]
            \addplot[only marks, mark=*, color=gray, opacity=0.5] table [x=weight_matched, y=dice, col sep=comma] {tables/abl_weight2.csv};
            \addplot[line width=2pt, color=gray, opacity=0.5, smooth] table [x=weight_matched, y={create col/linear regression={y=dice}}, col sep=comma] {tables/abl_weight2.csv};
            \end{axis}
            %\node at (current axis.above origin) [anchor=south, yshift=-4.1cm, xshift=2.25cm] {(b)};
        \end{tikzpicture}
        %\caption{}
        \label{fig:abl_wighting_b}
    \end{subfigure}
    \caption{Additional ablation on the introduced weighting term with the ACDC dataset. We find a different trend compared to Fig. 3, showcasing that the weight parameter must be tuned according to the dataset.}
    \label{fig:abl_wighting}
\end{figure}
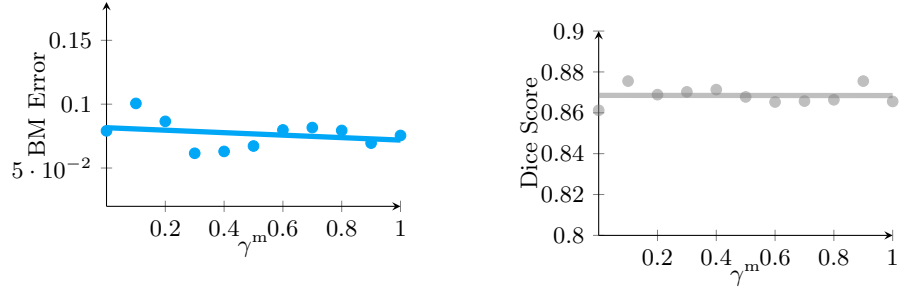

\begin{table}[h]
\centering
\label{tab:runtime}
\caption{Runtime comparison on two datasets for a single run.}
\begin{tabular}{lcc}
\toprule
\textbf{Loss}  & \textbf{OCTA-500} & \textbf{TopCoW}  \\
\midrule
Dice & 16m56s & 5m26s\\
ClDice & 17m17s & 7m32s\\
HuTopo & 59m27s & 19m01s\\
Ours & 28m10s & 10m39s\\
\bottomrule
\end{tabular}
\end{table}

\begin{figure}[h]
    \centering
    \includegraphics[width=0.8\linewidth]{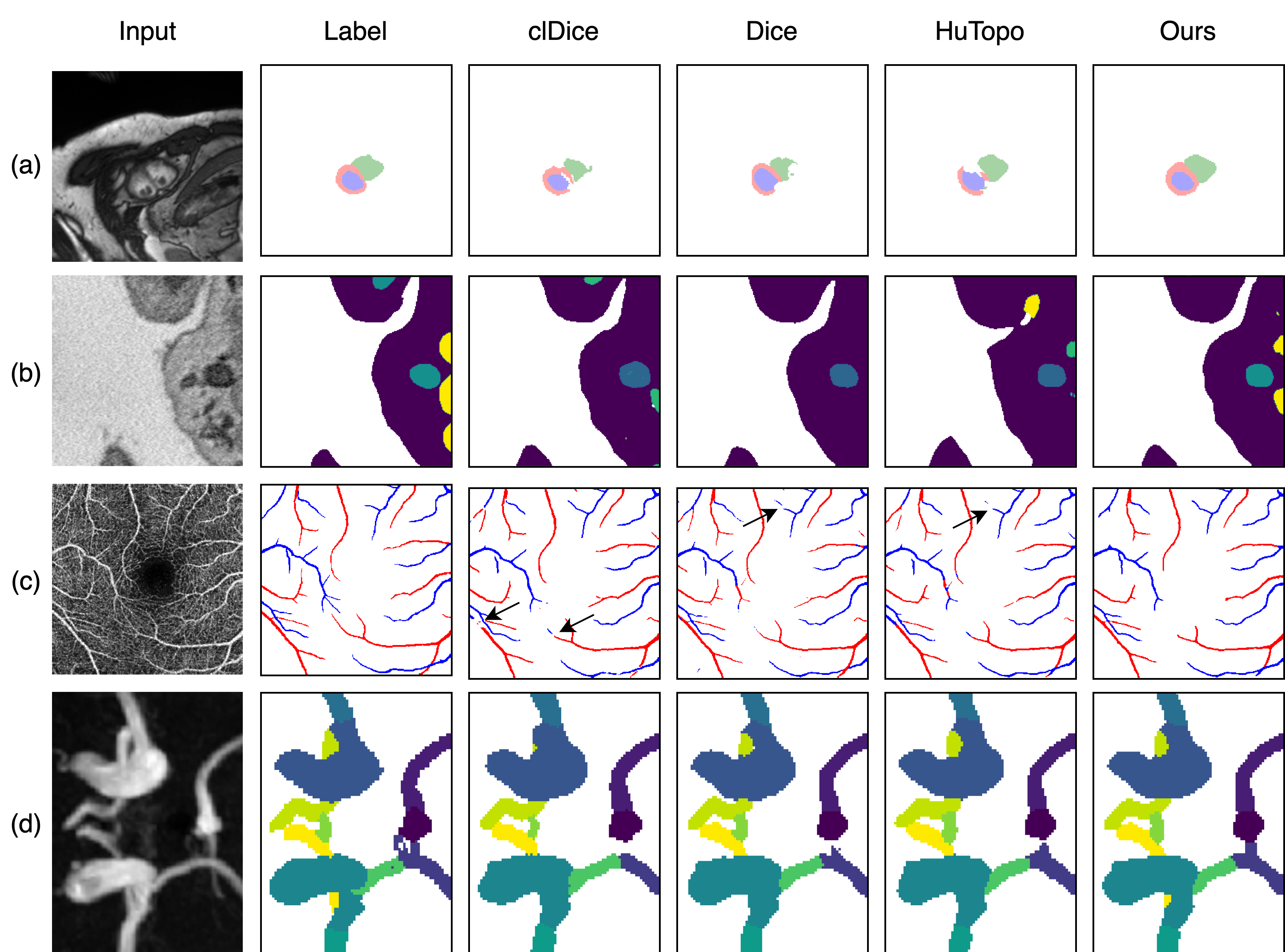}
    \caption{Additional qualitative results on ACDC (a), Platelet (b), OCTA-500 (c), and TopCoW (d) dataset.}
    \label{fig:suppl_qualitatives}
\end{figure}
\end{document}